\documentclass[aps,pre,preprint,showpacs,superscriptaddress,groupedaddress]{revtex4}     
\usepackage{latexsym}
\usepackage{natbib}
\usepackage{amsmath}
\usepackage{amsfonts}
\usepackage{amssymb}
\usepackage[T1]{fontenc}
\usepackage[dvips]{color,graphicx}   
\usepackage{verbatim}   
\usepackage{t1enc}
\usepackage{anysize}
\usepackage{float}  
\usepackage{natbib}
\usepackage{array}

\begin{document}

\widetext

\title{Slow epidemic extinction in populations with heterogeneous
  infection rates}

\author{C. Buono} 
\affiliation{Instituto de Investigaciones F\'sicas de Mar del Plata UNMdP-CONICET and Departamento de F\'sica FCEyN, Universidad Nacional de Mar del Plata, Funes 3350 (7600) Mar del Plata, Argentina.}
\author{F. Vazquez} 
\affiliation{Max-Planck-Institut f\"ur Physik
  Komplexer Systeme N\"othnitzer Str. 38, D-01187 Dresden, Germany.}
\affiliation{Instituto de F\'isica de L\'iquidos y Sistemas Biol\'ogicos UNLP-CONICET, Calle 59 Nro 789 (1900), La Plata
  Argentina.}  
\author{P. A. Macri}
 \affiliation{Instituto de Investigaciones F\'isicas
  de Mar del Plata (IFIMAR)-Departamento de F\'isica FCEyN-UNMDP-CONICET Funes 3350 (7600) Mar del Plata, Argentina.}
\author{L. A. Braunstein}
 \affiliation{Instituto de Investigaciones F\'isicas
  de Mar del Plata (IFIMAR)-Departamento de F\'isica FCEyN-UNMDP-CONICET Funes 3350 (7600) Mar del Plata, Argentina.}
\affiliation{Center for Polymer Studies, Boston University, Boston, Massachusetts 02215, USA.}

\begin{abstract}
We explore how heterogeneity in the intensity of interactions between
people affects epidemic spreading.  For that, we study the
susceptible-infected-susceptible model on a complex network, where a
link connecting individuals $i$ and $j$ is endowed with an infection
rate $\beta_{ij} = \lambda w_{ij}$ proportional to the intensity of
their contact $w_{ij}$, with a distribution $P(w_{ij})$ taken from
face-to-face experiments analyzed in Cattuto $et\;al.$ (PLoS ONE 5,
e11596, 2010).  We find an extremely slow decay of the fraction of
infected individuals, for a wide range of the control parameter
$\lambda$.  Using a distribution of width $a$ we identify two large
regions in the $a-\lambda$ space with anomalous behaviors, which are
reminiscent of rare region effects (Griffiths phases) found in models
with quenched disorder.  We show that the slow approach to extinction
is caused by isolated small groups of highly interacting individuals,
which keep epidemic alive for very long times.  A mean-field
approximation and a percolation approach capture with very good
accuracy the absorbing-active transition line for weak (small $a$) and
strong (large $a$) disorder, respectively.
\end{abstract}

\pacs{89.75.Hc,05.40.-a,87.19.X-,02.50.Ey}

\maketitle

\section{INTRODUCTION}

Nowadays, much of the daily human activity is constantly being
recorded by means of modern technologies, such as Internet, GPS,
mobile phones, blue-tooth and other electronic devices.  The gathering
and analysis of large amount of activity data have allowed the
exploration of statistical features of people's behavior.  In
particular, recent studies have revealed interesting properties about
how humans interact with each other, either by having conversations
\cite{Catt_01}, or sexual contacts \cite{Fox_06}, or by means of
mobile devices \cite{Karagiannis_07} or wireless communications
\cite{Scherrer_08}.  Despite the fact that the type of interactions in
these studies are quite different, they all found a common interaction
pattern, that is, human contacts are very heterogeneous.  This is
consistent with a broad distribution of different magnitudes that
quantify the timing of contacts, such as their duration, frequency and
gaps.  This diversity could eventually have an impact on propagation
processes that involve human contact, such as the spreading of rumors
or diseases.

In this article we explore epidemic spreading on a population with
heterogeneous interaction intensities.  We use a distribution of
intensities extracted from the pattern of contacts between
participants of a conference \cite{People_05}, obtained in recent
face-to-face experiments \cite{Catt_01}.  For the spreading process we
use the Susceptible-Infected-Susceptible (SIS) \cite{Bailey_75}
dynamics on Erd\H{o}s-R\'enyi (ER) networks \cite{Erd_01}, with
infection rates across links that are proportional to the intensity of
encounters.

As the rate of infection increases, the original SIS model \cite{Bailey_75}
exhibits a transition from an absorbing (disease-free) phase where
the infection dies exponentially fast to an active (endemic) phase
where the infection spreads over a large fraction of the population
and becomes persistent.  We find that the heterogeneity in the
intensity of contacts introduces an intermediate absorbing region, in
which the epidemic dies very slowly, as a stretched exponential or a
power law in time.  We experiment with other rate distributions and
show that this slow approach to epidemic extinction is caused by the
presence of small clusters composed by links with high infection
rates, which remain infected for very long times.  We also discuss
analogies with the effects observed in models with quenched disorder
\cite{Vojta_06}.

While our results are mainly concerned with the decay of the infection
in the epidemic-free phase, some related models
\cite{Stehle_11,Yang_12,Kars_01} have focused, instead, on the disease
prevalence within the endemic phase, or study the spreading power of a
given node \cite{Garas_10} using the susceptible-infected-recovered
dynamics.  Other studies have introduced heterogeneity at the
individual level, by assigning power law intertime events 
\cite{A_Vaz_01,Min_11}, node-dependent infection
rates \cite{Munoz_04}, or topology-dependent weight patterns
\cite{Odor_12,Odor_13,Odor_13-1}.  In our model heterogeneity is at the
interaction level, by means of link-dependent infection rates which
are not correlated with the topology of the network.

\section{SIS DINAMICS WITH FACE-TO-FACE DISORDER}

In the SIS model
\cite{Bailey_75}, each individual of a population can be either
susceptible (healthy) or infected.  Infected individuals transmit the
disease to its susceptible neighbors in the network at a rate $\nu$
and return to the susceptible state at a rate $\gamma$.  The dynamics
is controlled by the rescaled infection rate $\lambda = \nu/\gamma$.
For $\lambda$ above a critical value $\lambda_c$, even a small initial
fraction of infected nodes is able to propagate the disease through
the entire network (active phase), while for $\lambda < \lambda_c$ the
disease quickly dies out (absorbing phase), following an exponential
decay in the number of infected nodes.

\begin{figure}
\includegraphics[width=75mm]{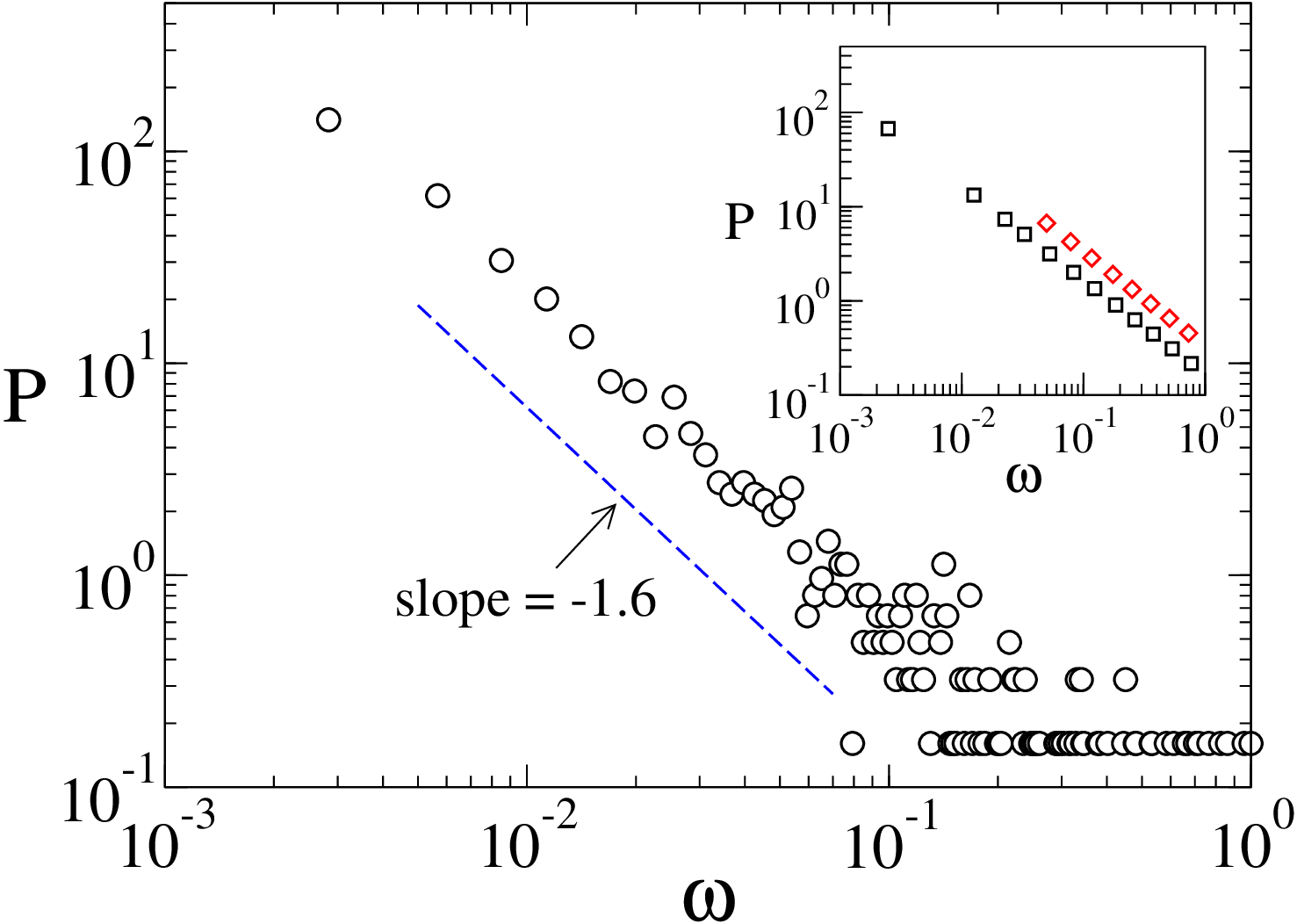}
\caption{Probability distribution of face-to-face contacts intensities
  (weights $w$) of the 25th Chaos Communication Conference in Berlin, on
  a log-log scale.  Intensity is defined as the total number of
  packages exchanged between two attendees, which is proportional to
  the contact duration.  The large dispersion of the data reflects the
  large heterogeneity in the duration of contacts.  Weights are
  rescaled to the interval $[0,1]$ for a better comparison with the
  theoretical distribution $P(w)=1/a w$ in the interval $[e^{-a},1]$,
  as shown in the inset for $a=6$ (squares) and $a=3$ (diamonds).}
\label{fig1}
\end{figure}

This model describes disease spreading in an ideal population where
transmission rates between individuals are all the same.  However, in
real populations we expect interactions to be heterogeneous, having a
broad range of intensities, as recently measured by analyzing mobiles
phone data \cite{Karagiannis_07} and by means of person-to-person
experiments \cite{Catt_01}.  In order to explore how the behavior of
the SIS model is affected by the heterogeneity of interactions, we run
simulations of the dynamics on ER networks with infection rates
distributed according to the weight distribution $P(w)$ of
face-to-face experiments \cite{Catt_01,People_05} (see
Fig.~\ref{fig1}).  In these experiments, participants of a three-day
conference were asked to wear a {\it radio frequency identification}
device on their chest, so that when two persons were close and facing
each other a relation of face-to-face proximity was registered.  The
weights $w$ of Fig.~\ref{fig1} are defined as the total number of
packets exchanged (or total contact time) between pairs of
participants during the three days.

We are assuming that infection rates are proportional to the total
time individuals are in contact with each other, as the likelihood of
transmission increases with exposure time -longer contacts imply a
higher risk of infection-.  Therefore, we assign an \emph{effective
  rate of infection} $\beta_{ij}=\lambda w_{ij}$ between two
individuals $i$ and $j$ that are connected by a link of weight
$w_{ij}$, where $\lambda$ is a free parameter that acts as a
transformation scale of contact intensities into infection rates.

\begin{figure}
 \includegraphics[width=75mm]{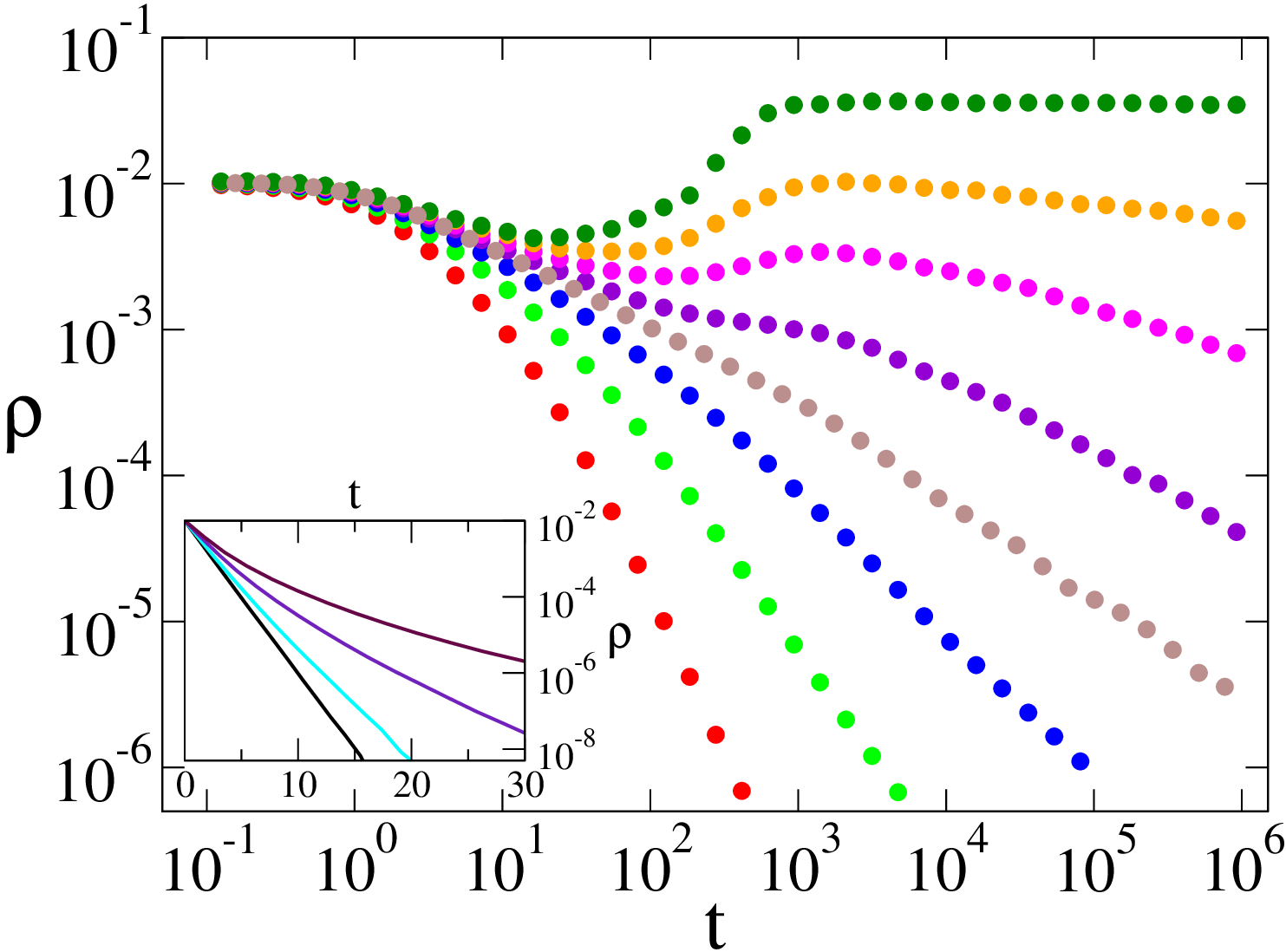}
 \caption{Average density of infected nodes $\rho$ vs time $t$, on a
   log-log scale, under the SIS dynamics on ER networks with $\langle
   k \rangle=4$ and $N=10^5$ nodes.  The infection rate distribution
   $P(\beta)$ corresponds to the weight distribution $P(w)$ of
   Fig.~\ref{fig1}, with $\beta = \lambda w$, for values of the
   control parameter $\lambda=3.9, 3.7, 3.6, 3.5, 3.4, 3.3, 3.0, 2.5$
   (main plot) and $\lambda=1.5, 1.0, 0.5, 0.2$ (inset), from top to
   bottom.  $\rho$ decays as a power-law for $2.0 \lesssim \lambda
   \lesssim 3.7$, as a stretched exponential for $0.25 \lesssim
   \lambda \lesssim 2.0$, and as an exponential for $\lambda \lesssim
   0.25$, as shown in the inset on a linear-log scale.}
\label{fig2}
\end{figure}

In Fig.~\ref{fig2} we show simulation results of the time evolution of
the average density of infected individuals $\rho$, over many
realizations of the SIS dynamics, starting from a configuration where
a small fraction of nodes have been randomly infected, and with
infection rates following the distribution of Fig.~\ref{fig1}.  All
simulations in this article correspond to ER networks of mean degree 
$\langle k \rangle=4$ and $N=10^5$ nodes.  We understand that ER
networks are an oversimplification of the complex topology of
interaction between attendees at the Berlin's conference, which is known
to have a broad degree distribution peaked at an intermediate value 
(similar to a Poissonian) as well as topological and temporal correlations
due to the intricate pattern of contacts \cite{isella_01}.  However, ER
networks, which have a Poisson degree  
distribution but are uncorrelated, are simple enough to allow for an 
exploration of the effects of the heterogeneity in the interaction
strengths, avoiding other possible effects due to its specific
degree distribution and correlations.

We found that, besides the typical behavior observed in the active and
exponential phases of the \emph{classic} SIS model (all infection rates
are the same), there is an intermediate region between $\lambda =
0.25$ and $3.7$ with very slow relaxation to the absorbing state.  The
region $2.0 \lesssim \lambda \lesssim 3.7$ is characterized by a
power-law decay with a continuously varying exponent, while in the
region $0.25 \lesssim \lambda \lesssim 2.0$ the decay is faster than a
power law but slower than exponential (see the inset of
Fig.~\ref{fig2}), and can be fitted by a stretched exponential.

\section{SIS dynamics with variable disorder strength}

In order to understand this phenomenon we explore the dynamics for
different distributions of weights.  We assign to each link $ij$ a
weight $w_{ij}= e^{-ar_{ij}}$, where $r_{ij}$ is a random number taken
from a uniform distribution in the interval $[0,1]$, and $a$ is a
parameter that sets the range of $w_{ij}$ in $[e^{-a},1]$.  This
method generates a power-law distribution $P(w)=1/a w$.  The parameter
$a$ controls the width of the distribution, and measures the
heterogeneity or strength of disorder.  In the inset of
Fig.~\ref{fig1} we plot $P(w)$ for $a=6$ and $a=3$, which is intended
to mimic the broad distribution of face-to-face contacts, even though
the decay exponents are different.

The distribution of infection rates is given by 
\begin{equation}
P(\beta)=\frac{1}{a \beta},~~~\mbox{with $\beta$ in $[\lambda
    e^{-a},\lambda ]$}.
\label{Pbeta}
\end{equation}
Notice that when $a \to 0$ we recover the classic model where
$\beta_{ij}=\lambda$ for all $ij$.  This kind of disorder was already
used in several works on complex networks \cite{bra02,bra03,Buo_01}.  We
expect that high-weight links facilitate the spreading of infections
in our model, while low-weight links hinder the spreading.

\begin{figure}[th]
 \includegraphics[width=75mm]{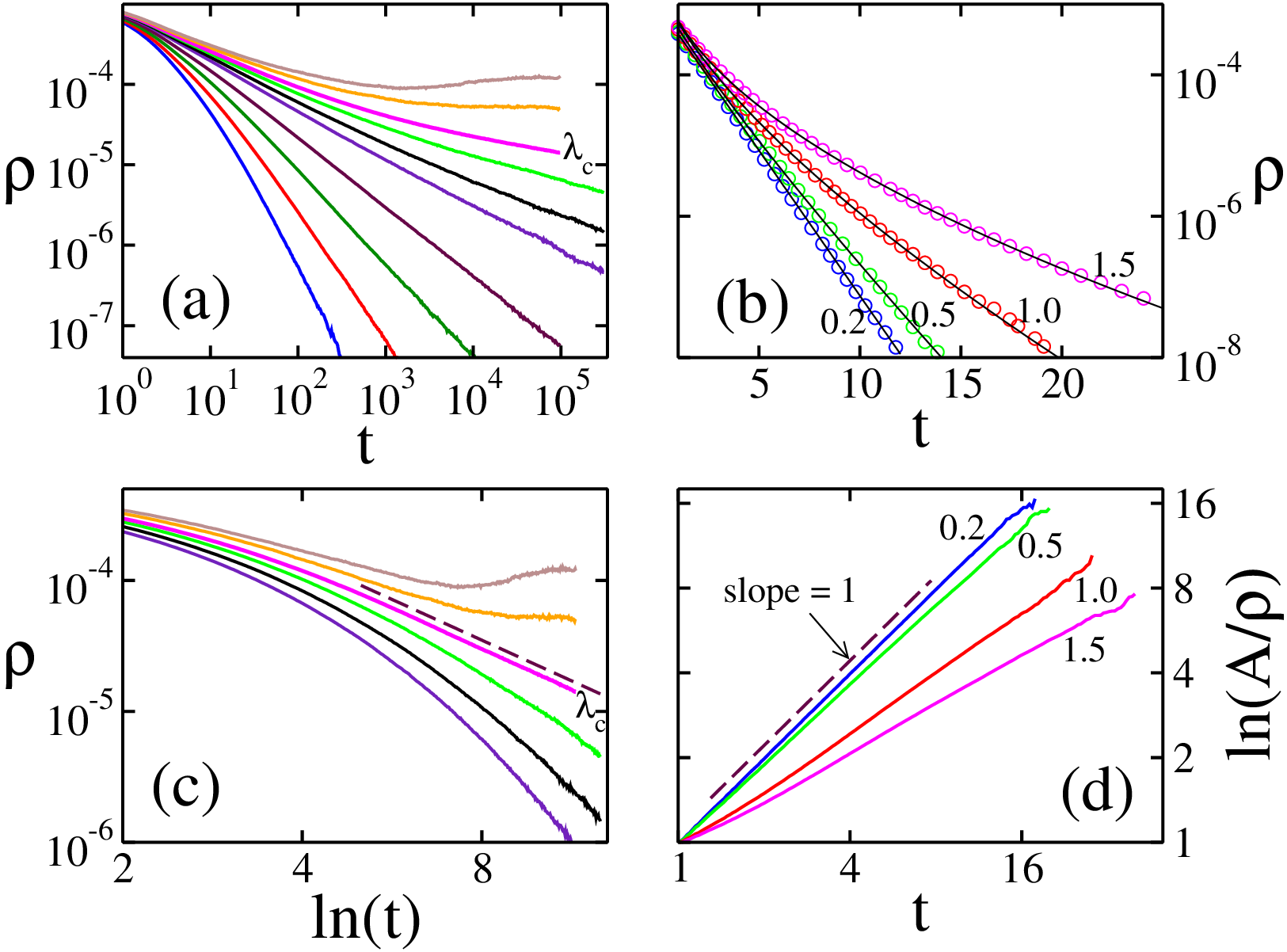}
 \caption{(Color online) SIS dynamics on ER networks with $\langle k
   \rangle = 4$, $N=10^5$ and distribution of infection rates
   $P(\beta)=1/a\,\beta$, with $a=20$ and $\beta$ in $[\lambda
     e^{-a},\lambda]$. (a) $\rho$ vs time on a double logarithmic
   scale, for $\lambda=6.0, 5.8, \lambda_c \simeq 5.56, 5.4, 5.2, 5.0,
   4.5, 4.0, 3.5$ and $3.0$ (from top to bottom).  $\rho$ decays as a
   power law for $2.0 \lesssim \lambda \lesssim \lambda_c$. (b) $\rho$
   vs time on a linear-log scale, for $\lambda=0.2, 0.5, 1.0$ and
   $1.5$ (circles).  Straight lines are best fittings using the
   function $A \,e^{-\alpha t^b}$ with $A=0.0009735$,
   $\alpha=0.94731$, $b=1.0$ for $\lambda=0.2$; $A=0.0016$,
   $\alpha=1.21$, $b=0.86$ for $\lambda=0.5$; $A=0.0028$,
   $\alpha=1.61$, $b=0.69$ for $\lambda= 1.0$; and $A=0.0045$,
   $\alpha=2.03$, $b=0.54$ for $\lambda=1.5$.  Only $\lambda=0.2$ is
   in the exponential region $\lambda<0.25$, while the other curves
   are stretched exponentials.  (c) $\rho$ vs $\ln t$ on a log-log
   scale, showing the extremely slow decay $\rho \sim (\ln
   t)^{-\beta}$ at the active-absorbing transition point $\lambda_c$.
   (d) The stretched exponential behavior for $\lambda = 0.5, 1.0$ and
   $1.5$ is shown as a straight line by plotting $\ln (A/\rho)$ vs
   time on a log-log scale.}
\label{fig3}
\end{figure}

\begin{figure}[th]
 \includegraphics[width=75mm]{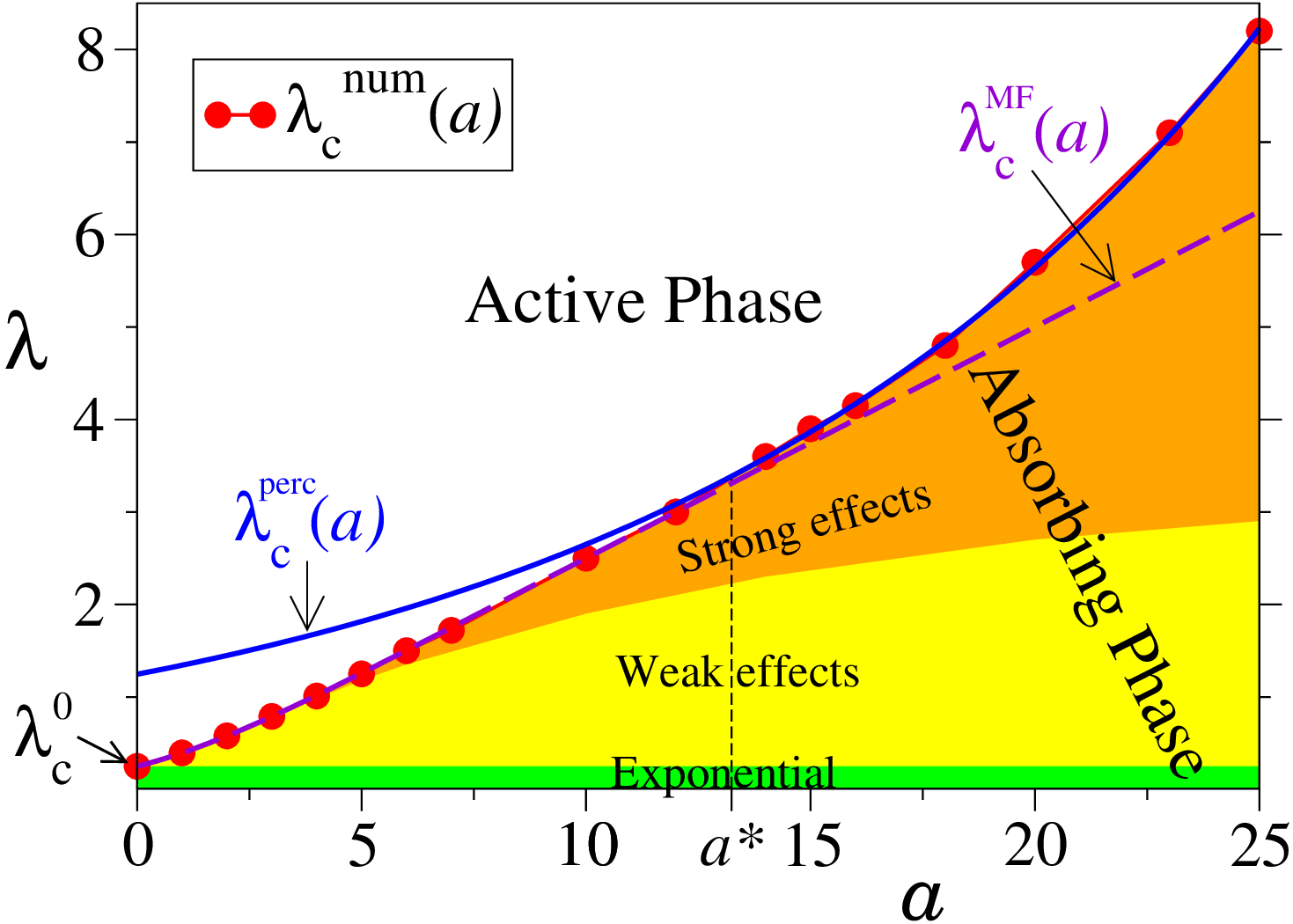}
 \caption{Phase diagram of the SIS model with infection rate
   distribution $P(\beta)=1/a\,\beta$, and $\beta$ in $[\lambda
     e^{-a},\lambda]$.  Colored regions correspond to the absorbing
   phase: orange and yellow for the strong and weak effects regions,
   and green for the exponential decay region.  The dashed and solid
   lines are the MF [Eq.~(\ref{lambdacMF})] and percolation
   [Eq.~(\ref{lambdacperc})] approximations, respectively, for the
   transition between the active and absorbing phases.}
\label{fig4}
\end{figure}

The behavior of $\rho$ under the theoretical disorder given by
Eq.~(\ref{Pbeta}) is very similar to the one observed in
Fig.~\ref{fig2} for the F2F disorder, showing a slow relaxation to the
absorbing state, as we can see in Figs.~\ref{fig3}(a) and
\ref{fig3}(b).  This suggests that the effect of disorder is quite
robust, since results seem to be independent on the power law exponent
of the distribution of weights.  In the $a-\lambda$ phase diagram of 
Fig.~\ref{fig4} we
summarize the different types of behaviors.  Above the numerical
transition line $\lambda_c^{\mbox{\tiny num}}(a)$ denoted by the red
circles we find the \emph{active phase} (white), where $\rho$ reaches
a stationary value larger than zero, and below we find the
\emph{absorbing phase} where $\rho$ decays to zero.  The transition
line corresponds to the value of $\lambda$ for which the decay is
algebraic in the logarithm of time, $\rho \sim (\ln t)^{-\beta}$
\cite{Vojta_06}, as shown in Fig.~\ref{fig3}(c).  The
absorbing phase is divided into three regions.  The \emph{exponential
  region} (green), which appears for $\lambda < \lambda_c^0$,
characterized by the decay $\rho \sim e^{-\alpha t}$ of the classic
model [see Figs.~\ref{fig3}(b) and \ref{fig3}(d)], the \emph{weak
  effects region} (yellow) where we observe an stretched exponential
behavior $\rho \sim A \,e^{-\alpha t^b}$ ($b<1.0$) [see
  Figs.~\ref{fig3}(b) and \ref{fig3}(d)], and the \emph{strong effects region} (orange),
with a power law decay $\rho \sim t^{-\gamma}$ [see
  Fig.~\ref{fig3}(a)].  Exponents $\alpha$, $b$ and $\gamma$ vary
continuously with $\lambda$ and $a$.  Along the line separating the
weak and strong effects regions, we observe a crossover between the
pure stretched exponential and power law decays.
 
\subsection{Active-absorbing transition line: mean-field and
  percolation approaches}  

In order to gain an insight about the dynamics of the model we
develop in this section a theoretical estimation of the transition
line between the active and absorbing phases of Fig.~\ref{fig4}. 
Within a mean-field (MF) approximation, $\rho$ evolves according to
$\dot{\rho}=-\rho+\lambda \langle w \rangle \langle k \rangle \rho
(1-\rho)$, where $\lambda \langle w \rangle = \lambda (1-e^{-a})/a$ is
the average infection rate, and $\langle k \rangle (1-\rho)$ is the
average number of susceptible neighbors of an infected node.  The
stationary solutions $\rho=0$ and $\rho=\left( \lambda -
\lambda_c^{\mbox{\tiny MF}} \right)/\lambda$ correspond to the
absorbing and active phases, respectively, with the transition point at
\begin{equation}
\lambda_{c}^{\mbox{\tiny MF}}(a)=\frac{a}{\langle k\rangle (1-e^{-a})},
\label{lambdacMF}
\end{equation}
For $a=0$ we recover the \emph{classic} transition point 
$\lambda_c^0 \equiv \lambda_c(a=0)=1/\langle k \rangle = 0.25$ of the
classic model.  \emph{Impurities}, in the form of low-weight links, locally reduce
infection rates, thus the transition happens at a value
$\lambda_{c}^{\mbox{\tiny MF}}(a) > \lambda_{c}^0$.

Expression~(\ref{lambdacMF}) (dashed line in Fig.~\ref{fig4}) is a
very good estimate of $\lambda_c^{\mbox{\tiny num}}(a)$ for $a\lesssim
14$ (weak disorder), but systematic deviations appear as $a$
increases.  Discrepancies arise because MF assumes that \emph{all}
links can spread the disease but, when $a$ is large, a fraction of
links have such small rates (inactive links) that infection never
passes through them during the epidemic's life time, and thus the
\emph{effective} network for the spreading dynamics is diluted with
respect to the original network.  When dilution is large enough the
effective network gets fragmented into many small disconnected
components and, as the disease cannot spread out of these components
the active state is never reached.  Therefore, the active-absorbing
transition point for $a$ large (strong disorder) corresponds to the
\emph{percolation threshold}.  This occurs when the fraction of
inactive nodes $q$ (nodes attached only to inactive links) exceeds the
critical value $q_c$.  For ER networks with Poisson degree
distribution $P_k=e^{-\langle k\rangle} \frac{\langle k \rangle
  ^{k}}{k!}$ is
\begin{eqnarray}
q = \sum_k l_{\mbox{\tiny I}}^k P_k = e^{-\langle k \rangle
  (1-l_{\mbox{\tiny I}})},~~~~\mbox{where} \\ l_{\mbox{\tiny I}}
\equiv \int_{\lambda e^{-a}}^{\beta_{m}} P(\beta) \,d\beta =
\frac{1}{a} \ln\left(\frac{\beta_{m} e^a}{\lambda} \right)
\label{eq5}
\end{eqnarray}
is the fraction of inactive links, and $\beta_m$ is the largest
infection rate that does not allow the transmission of the disease.
At the percolation threshold $\langle k\rangle=(1-q_c)^{-1}$ in the
$N\to\infty$ limit, thus $q_c = \exp \Big\{(1-q_c)^{-1} \left[ \ln
  \left( \beta_m \,e^a / \lambda_c^{\mbox{\tiny perc}} \right) /a -1
  \right] \Big\}$, from where the percolation transition line is
\begin{equation}
\lambda_c^{\mbox{\tiny perc}}(a) = \beta_m  \; q_c ^{-a (1-q_c)}.
\label{lambdacperc}
\end{equation}
Using $q_c=0.7443$ for a network of size $N=10^5$ \cite{Wu_01},
expression~(\ref{lambdacperc}) with $\beta_{m} \simeq 1.2457$ is in
excellent agreement with $\lambda_c^{\mbox{\tiny num}}(a)$ for $a
\gtrsim 14$ (solid line in Fig.~\ref{fig4}).  The value of $\beta_m$
is estimated from the crossover conditions $\lambda_{c}^{\mbox{ \tiny
    MF}}(a)=\lambda_c^{\mbox{\tiny perc}}(a)$ and $\partial
\lambda_{c}^{\mbox{ \tiny MF}}(a)/\partial a=\partial
\lambda_{c}^{\mbox{ \tiny perc}}(a)/\partial a$ between the MF and
percolation lines at the weak-strong disorder crossing point $a=a^*$.
We obtain $\beta_{m}=-[e \ln q_c]^{-1}$ and $a^*=-[(1-q_c) \ln q_c
]^{-1}$ with $a^* \simeq 13.243$ for the network used here.

In the next section we analyze in more detail the dynamics in the
absorbing phase and provide an explanation about the origin of slow
relaxations.

\subsection{Anomalous behavior in the absorbing phase}

\begin{figure}
 \includegraphics[width=75mm]{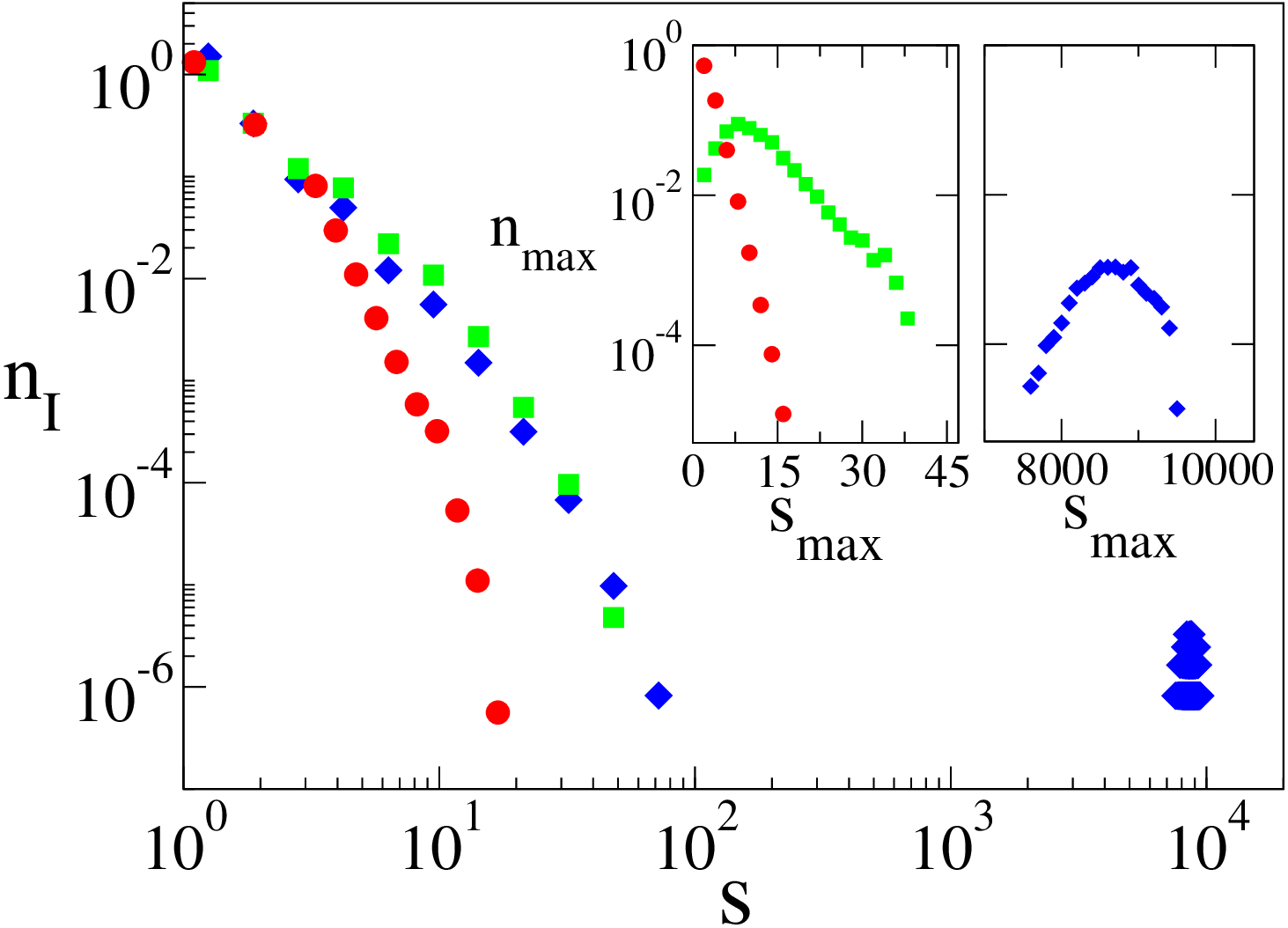}
 \caption{Cluster size distribution of infected nodes $n_I$ for
   $a=6$ and values of $\lambda$ inside three different regions of
   Fig.~\ref{fig4}: $\lambda=0.75$ (circles) and $1.4$ (squares) in
   the weak and strong effects regions, and $2.0$ (diamonds) in the
   active phase.  Distributions correspond to snapshots of the network
   at fixed large times.  The average infection rates inside each
   cluster are $\langle \beta \rangle \simeq 0.27$, $0.70$ and $0.75$
   for $\lambda=0.75$, $1.4$ and $2.0$, respectively.  Insets: size
   distribution of the largest cluster $n_{\mbox{\tiny max}}$ showing
   the appearance of a large component in the active phase.}
\label{fig5}
\end{figure}

We have seen that the heterogeneity in infection rates induces a large new
region inside the absorbing phase, in which the temporal evolution of
$\rho$ exhibits an anomalous slow decay.  This is caused by the
presence of exponentially small isolated regions in the network where
the system is locally active, that is, with infection rates
$\beta_{ij} > \lambda_c^0$, which are able to sustain the activity for
very long times.  To check this, we calculated the size distribution
of clusters composed only by infected nodes $n_I(s)$, at a fixed large
time.  Results are shown in Fig.~\ref{fig5} for $a=6$.  Inside the weak and
strong effects regions ($\lambda=0.75$ and $1.4$), $n_I(s)$ is close
to an exponential, and the size of the largest cluster $s_{\mbox{\tiny
    max}}$ is much smaller than the network size $N=10^5$ (see inset
of Fig.~\ref{fig5}).  Also, the values $0.27$ and $0.70$ of the
average infection rates inside these clusters for $\lambda=0.75$ and
$1.4$, respectively, show that the long-time activity is located
inside \emph{active clusters}, in which the average rate of infection
$\langle \beta \rangle > \lambda_c^0$.  For comparison, in the active
phase ($\lambda=2.0$) is $7600 \lesssim s_{{\mbox \tiny max}} \lesssim
9500$, indicating the spreading of the disease over a large fraction
of the network.

Similar anomalous behaviors are found in models with disorder, giving
rise to the so-called Griffiths phases (GP)
\cite{Vojta_06,Munoz_04,Odor_12,Odor_13,Odor_13-1,Vaz_03,Mar_12}.  The
combination of exponentially rare regions in space that survive for
exponentially long times results in an overall slowing down of the
dynamics, as we show below.  The long-time contribution of active
clusters to $\rho$ is estimated as
\begin{equation}
\rho \sim \int ds \, s \, P(s) \, e^{-t/\tau(s)},
\end{equation} 
where $P(s) \sim e^{-\tilde p s}$ \cite{New_03} is the fraction of
active clusters of size $s$ and $\tau(s)$ is the mean decay time of
those clusters.  By doing a saddle-point analysis, and using the
finite-size scaling $\tau(s) \sim e^{c s}$, one arrives to the
power-law decay $\rho \sim t^{-\tilde p/c}$ (with logarithmic
corrections) observed in the strong effects region of Fig.~\ref{fig4}.
The size of active clusters is of order one for $\lambda$ just above
$\lambda_c^0$, leading to exponentially weak effects of the form $\rho
\sim e^{-\alpha t^b}$ \cite{Vojta_06}.

\section{Discussion and Conclusions}  

In summary, the heterogeneity in the intensity of contacts between
individuals induces a regime with extremely slow (power-law or
stretched exponential) relaxation to epidemic extinction, akin to the
slowing down found in systems with quenched disorder.  This effect is
very robust, as it was observed using an empirical distribution of
contact durations in face-to-face experiments, as well as a
theoretical distribution with variable width.  Given that both are
long tailed distributions but with different exponents, we suspect
that the anomalous relaxation is observed in general for broad weight
distributions.  To check this concept we run simulations (not shown)
using a bimodal distribution
$P(\beta)=p\,\delta(\beta-\beta_1)+(1-p)\,\delta(\beta-\beta_2)$, with
$0 \le p \le 1$ and $\beta_2 > \beta_1$ \cite{Vojta_06}.  We observed
slow relaxations for $\beta_2 > \lambda_c^0 > \beta_1$, that is, when
there are finite fractions of links with infection rates above and
below the classic transition point $\lambda_c^0$.

In order to explore whether these effects depend on the specific
topology of interactions,  we have done some testings with scale-free
networks.   We found that the active-absorbing transition line on the
phase diagram of Fig.~\ref{fig4} is shifted down to very small values, but we
could not clearly identify a finite region with slow decay.
Therefore, we suspect that rare-region effects are not present in
networks with heterogeneous degree distributions.  This is probably
because weights are randomly distributed over the network, thus
high-degree nodes always spread the disease (it is very unlikey that
all links attached to hubs have very low weights).  Instead, assigning
weights according to the topology of the network may induce
rare-region effects, as it was shown in
\cite{Odor_12,Odor_13,Odor_13-1} using Barabasi-Albert trees with
disassortative weighting.  It would be worthwhile to perform a deeper
analysis to study how relaxations are affected by other properties of real
contact networks, such as topological and temporal correlations. 
 
While \emph{temporal} heterogeneity, causality and bursty activity was
found to hinder spreading \cite{Kars_01,Min_11}, we showed here that
\emph{spatial} heterogeneity has the counterbalanced effect, making
the epidemic more persistent by slowing down its extinction.  Once a
group of highly interacting individuals gets infected, they are able
to continuously reinfect each other at a high rate, keeping the
infection inside the group for very long times.  Our findings can be
used to design efficient mitigation strategies for the disease.  For
instance, moderating the activity of highly interacting people could
dramatically speed up the final stage of the epidemic.

This work was financially supported by UNMdP and FONCyT (Pict
0293/2008). The authors thank Lucas D. Valdez for useful comments and
discussions.

\bibliography{Buono.bib}

\end{document}